\documentclass{optica-article}

\journal{opticajournal} % for journals or Optica Open

\articletype{Research Article}

\usepackage{lineno}
\usepackage{amsmath}
\usepackage{xcolor}% color text

\usepackage{graphicx}

\usepackage{mathrsfs}
\usepackage{multicol}
\usepackage{notoccite}

\newcommand{\ud}{\mathrm{d}}

\graphicspath{ {./figures/} {./}}

\begin{document}

%\title{Modulational instability in telegraph random dispersion-managed fiber links}
%\title{Random telegraph dispersion in optical fibers: modulational instability}
\title{Random telegraph dispersion-management: modulational instability}

\author{Andrea Armaroli\authormark{1,2,*} and Matteo Conforti\authormark{1}}

\address{\authormark{1}Univ.~Lille,  CNRS,  UMR  8523-PhLAM-Physique  des  Lasers  Atomes  et  Mol\'ecules,  F-59000  Lille,  France\\
\authormark{2}Universit{\`a} di Ferrara, Department of Engineering, via Saragat 1, I-44122, Ferrara, Italy}

\email{\authormark{*}andrea.armaroli@unife.it} %% email address is required; see note below about the corresponding author designation

% use {asbstract*} to suppress the copyright line. Copyright information will be added in production

\begin{abstract*} 
We study modulational instability in a fiber system resembling a dispersion-managed link   where the sign of the group-velocity dispersion varies randomly according to a 
 telegraph process.
We find that the  instability gain of stochastic origin converges, for long fiber segment mean length (the inverse of the transition rate between the two values), to the conventional values found in a homogeneous anomalous dispersion fiber. {For short fiber segments, the gain bands are broadened and the maximum gain decreases.}
By employing  correlation splitting formulas, we obtain closed form equations that allow us to estimate the instability gain from the linearized nonlinear Schr{\"o}dinger equation. We compare the analytical to the numerical results obtained in a Monte Carlo spirit. The analysis is proven to be correct not only for a fluctuating group-velocity dispersion, but also including fourth-order dispersion (both constant or varying according to a synchronous or independent telegraph process).
These results may allow researchers to tailor and control  modulational instability sidebands, with applications in telecommunications and parametric photon sources. 

\end{abstract*}

%%%%%%%%%%%%%%%%%%%%%%%%%%  body  %%%%%%%%%%%%%%%%%%%%%%%%%%
\section{Introduction}
 Modulational instability (MI), \textit{i.e.}, the destabilization of a uniform wavepacket by exponentially growing harmonic perturbations around its carrier frequency, is an ubiquitous phenomenon in nonlinear dispersive wave physics\cite{Zakharov2009}. The first studies emerged in electromagnetic waves \cite{Bespalov1966} and hydrodynamics \cite{Benjamin1967a, Zakharov1968}, and about two decades later the phenomenon was observed in optical fibers \cite{Tai1986}. The basic ingredients that yield MI in a one dimensional homogeneous system are focusing cubic nonlinearity (like the Kerr effect in silica) and anomalous (negative) group-velocity dispersion (GVD).

 Nevertheless, MI exists also in normal GVD, provided that high-order dispersion \cite{Cavalcanti1991} or birefringence \cite{Biancalana2004e} are considered. Moreover, MI is observed in single-mode fibers in the normal GVD region also if the GVD is varied along the propagation direction. If the variation is periodic, the MI is equivalent to the destabilization of a harmonic oscillator subject to parametric forcing and is denoted as parametric MI\cite{Smith1996,Abdullaev1996,Droques2012,Armaroli2012,Mussot2018}. A resonance is found for every  integer order to correspond to a MI sideband: only a few of them are observable and their frequency separation from the carrier scales as the square root of their order. 
 
Random variations of GVD were also the subject of extensive study. In the late 90s, the white noise process (an exactly solvable model) was considered\cite{Abdullaev1996,Abdullaev1999,Garnier2000,Chertkov2001}. Only recently  different random processes were considered: localized GVD kicks \cite{Dujardin2021} or coloured processes of low-pass or band-pass type\cite{Armaroli2022}.

Up to this point, we mentioned only examples of {(possibly large)} GVD variations around a nonzero average GVD. Actually, to prevent the competition with conventional MI, which usually exhibits larger gain, normal average GVD is considered. %Large fluctuations are not forbidden, though.

Conversely, systems with zero- or nearly zero-average GVD have attracted the attention of many researchers, because dispersion-induced pulse-broadening is largely suppressed and nonlinear propagation optimized \cite{Agrawal2012,Turitsyn2012}. This strategy is usually denoted dispersion management (DM) and consists in alternating positive and negative GVD segments along the propagation direction. A uniform distribution for fluctuation of segment length is studied in \cite{Malomed2001,Armaroli2023DM} for pulse propagation and MI, respectively. 

Here we adapt part of the theory presented in Ref.~\cite{Armaroli2022} to  DM-like random fluctuations modelled as a telegraph process, \textit{i.e.}, a low-pass colored random process taking only two values alternating at exponentially distributed distances. Thanks to correlation splitting formulas\cite{Furutsu1963,Novikov1965,Shapiro1978,GittermanNoisy,Gitterman2005,Klyatskin2010book}, the MI equations (i.e. the linearized uni-directional propagation equation) are also  solvable. 

% COMMENT: In reality, each fiber span has big GVD, only the average is close to zero, but the fluctuactions are big
%For the operation point may be close to the zero-dispersion wavelength (ZDW), we explore the effect of the first higher-order effect having an impact on MI, \textit{i.e.}, the fourth-order dispersion (FOD).

{Our theoretical approach is quite general and flexible, and permits to account for additional physical effects. As a relevant example, we explore the influence of the first higher-order effect having an impact on MI, \textit{i.e.}, the fourth-order dispersion (FOD).}

The rest of the paper is organized as follows.
After presenting the model equations and detailing the analytical approach for fluctuations of GVD and (possibly) FOD in Sec.~\ref{sec:model}, we compare the analytical estimates to numerical results obtained in a Monte Carlo spirit in Sec.~\ref{sec:results}. Several different examples are discussed, including no FOD, fixed or fluctuating FOD. 
Conclusions are reported in Sec.~\ref{sec:concl}.

\section{Analytical approach}
\label{sec:model}
\begin{figure}[ht]
    \centering
    \includegraphics[width=0.5 \textwidth]{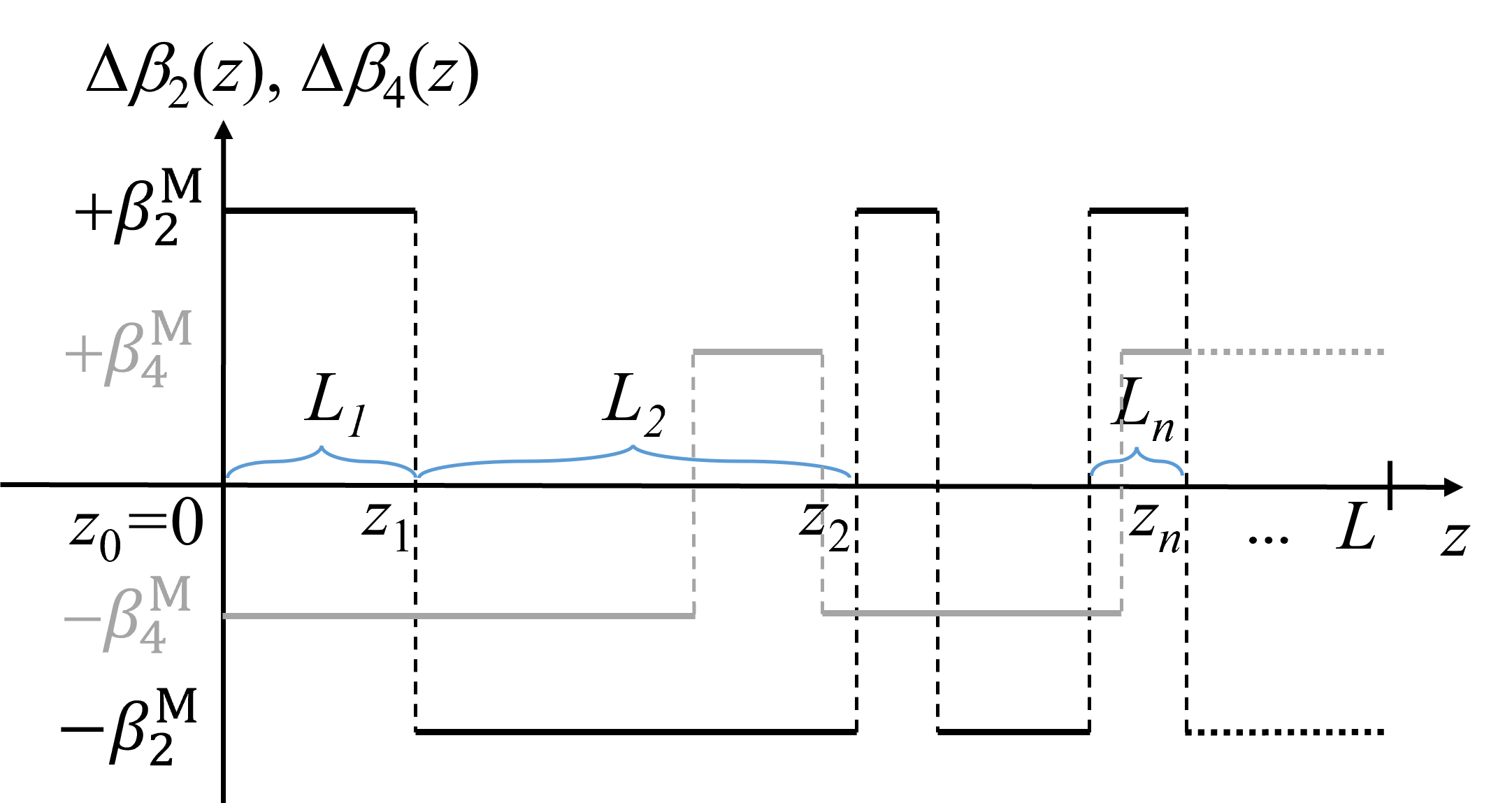}
    \caption{Schematic representation of the GVD and FOD profiles in a
typical fiber realization. The random distances of sign changes are shown only for GVD. }
    \label{fig:sketch}
\end{figure}
We consider the propagation of  optical pulses ruled by the generalized nonlinear Schr{\"o}dinger equation (GNLSE)\cite{Agrawal2012},
\begin{equation}
   i\partial_z U - \frac{1}{2}\beta_2(z)\partial_{t}^2U +  \frac{1}{24}\beta_4(z)\partial_{t}^2U + \gamma |U|^2U = 0,
\label{eq:NLS1}
\end{equation}
where $U(t,z)$ is the complex envelope of the optical field, ($t$,$z$) are time and propagation distance in a frame moving at the group velocity of the fiber mode, $\gamma$ the (constant) nonlinear coefficient,  $\beta_2(z) = \bar\beta_2 + \Delta\!\beta_2(z)$ is the  GVD, and $\beta_4(z) = \bar\beta_4 + \Delta\!\beta_4(z)$ is the FOD.  Moreover,    $\Delta\!\beta_2$ and $\Delta\!\beta_4$ are two  telegraph processes. {They can take only two values, $\pm{ \beta_2^\mathrm{M}}$, with ${ \beta_2^\mathrm{M}}>0$, (resp.~$\pm{ \beta_4^\mathrm{M}}$, with ${ \beta_4^\mathrm{M}}\ge 0$). We consider either (i) synchronous or (ii) independent (asynchronous) processes. The latter case is schematically} illustrated in Fig.~\ref{fig:sketch}. 
% Sign changes occur at $z_1, z_2, \ldots, L$, where $z_{n} = z_{n-1} +  L_n$, $n=1,2,\ldots$ and $z_0=0$.  The lengths $L_n$ are independent, identically distributed random variables of exponential probability distribution function with mean $\bar L_2$ (resp.~$\bar L_4$). 

{The samples of the dispersion profiles (GVD and FOD) are generated using the following procedure \cite{karlin}. The total length $L$ is fixed, then we extract for the $i$-th sample a positive integer $N_i$ from a Poisson distribution with rate $L/\bar L_m$, $m=2,4$. We extract $N_i$ points $w_n$, $n=1,2,\ldots,N_i$ from a uniform distribution in $[0,L]$, and order them in ascending order, to obtain the points $z_i$ where the sign changes occurs (in the case of asynchronous GVD and FOD two independent sets of points are used).  The lengths $L_n=z_{n} - z_{n-1}$ are independent, identically distributed random variables of exponential probability distribution function with mean $\bar L_m$, $m=2,4$, thus representing a telegraph process.}
%
%The total length $L$ is fixed, thus we extract for the $i$-th sample a positive integer $N_i$ from a Poisson distribution with rate $L/\bar L_m$, $m=2,4$ and then fix the points $z_n$, $n=1,2,\ldots,N_i$ from a uniform distribution in $[0,L]$. Equivalently, let Sign changes occur at $z_1, z_2, \ldots, z_{N_i}\le L$, where $z_{n} = z_{n-1} +  L_n$, $n=1,2,\ldots$ and $z_0=0$.  The lengths $L_n$ are independent, identically distributed random variables of exponential probability distribution function with mean $\bar L_m$, $m=2,4$.
The telegraph process is an example of lowpass colored noise, as discussed in detail in \cite{Armaroli2022}. The bandwidth is inversely proportional to the mean length, $B_m = \frac{2}{\bar L_m}$. 

A  continuous wave ($t$-independent) solution of Eq.~\eqref{eq:NLS1} reads  $U_0(z) = \sqrt{P}\exp(i \gamma P z)$, where $P$ is the carrier power. 
In order to study its stability, we insert in  Eq.~\eqref{eq:NLS1} the \textit{Ansatz}   ${U}(z,t) = \left[\sqrt{P} + \check x_1(z,t) + i \check x_2(z,t)\right]\exp (i \gamma P z)$, where $\check x_{1,2}$ are assumed to be small, linearize and Fourier-transform the resulting equation with respect to $t$ ($\omega$ is used as the associated angular frequency detuning from the carrier $U_0$).  We obtain  
 \begin{equation}
   \frac{\ud x}{\ud z}
	 	 = 
	 	\begin{bmatrix}
	 		0 & -g(z) \\h(z) & 0
	 	\end{bmatrix}
	 	x,
\label{eq:MIeqs1}
\end{equation}
with $x  \equiv ( x_1,x_2)^\mathrm{T}${\color{black}---$x_{1,2}$ are the Fourier transforms of $\check x_{1,2}$, functions of $\omega$ and $z$}, $g(z) = \beta_2(z) \frac{\omega^2}{2} + \beta_4(z) \frac{\omega^4}{24} =g_0 + \delta g(z)$ and $h(z) = g(z) +  2\gamma P = h_0 + \delta\! g(z)$, with $g_0\equiv \bar \beta_2 \frac{\omega^2}{2} + \bar\beta_4 \frac{\omega^4}{24}$, $h_0 \equiv g_0 + 2\gamma P$, $\delta\!g  \equiv \Delta\!\beta_2(z) \frac{\omega^2}{2} + \Delta\!\beta_4(z) \frac{\omega^4}{24}$.
Eq.~\eqref{eq:MIeqs1} is a system of stochastic differential equations (SDEs) for each value $\omega$.

By letting $ { \beta_2^\mathrm{M}}=  { \beta_4^\mathrm{M}}=0$, Eq.~\eqref{eq:MIeqs1} is reduced to a system of linear autonomous ordinary differential equations (ODEs) \cite{Cavalcanti1991,Pitois2003}. Provided that %$\bar\beta_2\bar\beta_4<0$ \textcolor{red}{or $\bar\beta_2,\bar\beta_4<0$}, 
{$\bar\beta_2$ and $\bar\beta_4$ are not both positive,}
there exist intervals in $\omega$ for which the eigenvalues of the matrix in Eq.~\eqref{eq:MIeqs1} are real and the MI gain thus reads
    \begin{equation} 
       G(\omega) = \max\left[\operatorname{Re}\left\{
           \sqrt{-\left(\bar\beta_2 \frac{\omega^2}{2} + \bar\beta_4 \frac{\omega^4}{24}\right)\left(\bar\beta_2 \frac{\omega^2}{2} + \bar\beta_4 \frac{\omega^4}{24}+2\gamma P \right)}\right\}\right].
    \label{eq:MIGNLSE}
 \end{equation}
It is easy to verify (directly or by nonlinear phase-matching arguments) that 
this expression has maxima in 
\begin{equation}
\omega^2 = -\frac{6\bar\beta_2}{\bar\beta_4}\left(1\pm \sqrt{1-\frac{2\gamma P}{3}\frac{\bar\beta_4}{\bar\beta_2^2}}\right),
    \label{eq:MIpeaks}
\end{equation}
{(provided that $\omega^2>0$ )} at which, remarkably, $G$ achieves the same value, $G_\mathrm{max}=\gamma P$, obtained in conventional MI. There exist three possible combinations of signs of $\bar\beta_2$ and $\bar\beta_4$ which lead to MI gain.
{For anomalous GVD and $\bar\beta_4<0$ a single sidelobe appears, which is a deformation of the conventional MI gain. If, instead,  $\bar\beta_4>0$ and $\gamma P\frac{\bar\beta_4}{\bar\beta_2^2}\ll 1$, two MI sidelobes appear: the first is a slight variation of the conventional MI sidelobe in the absence of FOD, \textit{i.e.}, $0\le\omega\le 2\sqrt{\frac{\gamma P}{\bar\beta_2}}$ and the second occurs at $\omega\approx \sqrt{12\left|\frac{\bar\beta_2}{\bar\beta_4}\right|}$. Around this same value a single MI sidelobe appears for  normal GVD and $\bar\beta_4<0$. } For the sake of precision, we 
notice that the high frequency narrow MI sidebands are slightly different in anomalous vs.~normal GVD for the same $|\bar\beta_4|$, as it is apparent from Eq.~\eqref{eq:MIpeaks}.
Equations \eqref{eq:MIGNLSE} and \eqref{eq:MIpeaks} will be important as limits of the analytical estimates presented below.
 
In the following we assume {vanishing average GVD} $\bar\beta_2=0$ (unless we refer to conventional MI, where $\bar\beta_2 = -1$);  {the average FOD} $\bar\beta_4$ can take any positive or negative value.

In \cite{VanKampenBook} a precise discussion of multiplicative noise (the fluctuation of the natural frequency of a harmonic oscillator, assumed non null) is presented. Its application to Eq.~\eqref{eq:MIeqs1} was discussed in \cite{Armaroli2022}: the unperturbed system is stable and the first moments $\langle x \rangle$ cannot provide any hint {on the stability of the perturbed system}. 
The equations for the second moments of the distribution of $x$, which are directly related to the fiber output power,  need studying. 
%{\color{green} Nun me piace, l'amma a cagn{\'a}} In \cite{Armaroli2023DM} we presented \sout{instead} an example of competing deterministic and stochastic effects. Consider the limit of $\bar\beta_4=0$:the average wavenumber is $k\equiv \sqrt{g_0h_0}=0$, the oscillator is at the margin of stability for all $\omega$, subject to a random alternation between stable and unstable regimes. The first moment can be non-negative for some values of $\omega$. The discussion in \cite{Armaroli2023DM} showed that the first moments express the residual deterministic MI gain, but do not correspond to a physically measurable quantity; instead, the second moments are directly related to the fiber output power.

We let $X_1 = x_1^2$, $X_2 = x_2^2$, and $X_3 = x_1x_2$ and derive from Eq.~\eqref{eq:MIeqs1}
\begin{equation}
	 	 	\frac{\ud}{\ud z}
	 	 {X} = 
	 	\begin{bmatrix}
	 		0 & 0 & -2 g(z) \\
	 		0 & 0 & 2 h(z) \\
	 		h(z) & -g(z) & 0
	 	\end{bmatrix}
	 	 {X},
	 	\label{eq:MIeqs2}
\end{equation}
with $X\equiv (X_1,X_2,X_3)^\mathrm{T}$.
The following discussion relies only on Eq.~\eqref{eq:MIeqs2}. We first study the effect of fluctuations of GVD (with fixed average FOD), next we study the fluctuations of both GVD and FOD. 

\subsection{Fluctuations of GVD, ${\beta_4^\mathrm{M}}=0$}
\label{sec:singleprocessGVD}

{Let us consider a constant FOD ($\Delta\!\beta_4=0$)}, therefore $\delta\!g = \Delta\! \beta_2(z) \frac{\omega^2}{2}$. 
The telegraph process allows for a particularly simple closure of the moment equations\cite{Gitterman2005,GittermanNoisy}. We apply to Eq.~\eqref{eq:MIeqs2} the Shapiro-Loginov formula \cite{Shapiro1978} 
\begin{equation}
    \langle\Delta\!\beta_2 \frac{\ud X_i}{\ud z}\rangle = \left(\frac{\ud}{\ud z} + B_2\right)\langle\Delta\!\beta_2 X_i\rangle.
    \label{eq:ShapiroLoginov}
\end{equation}

Two steps are required: (i) directly average Eq.~\eqref{eq:MIeqs2}, (ii) multiply each row by $\Delta\beta_2$ and average. Three auxiliary variables are introduced $X_{3+i} \equiv \Delta\beta_2 X_i$,  for  $i=1,2,3$. Taking into account  the property of telegraph processes, $\langle\Delta\!\beta^2_2 X_i\rangle = \sigma^2_2\langle X_i\rangle$, with $\sigma^2_2\equiv ({\beta_2^\mathrm{M}})^2$, 
we obtain a sixth order system of ODEs,
\begin{equation}
  \frac{\ud \langle X\rangle}{\ud z}
	 	= \left[
 	 	     \begin{array}{c|c}
 	 	     A_4 &  C_2 \\
 	 	    \hline
 	 	    \sigma^2_2  C_2 &  A_4-B_2 \mathbf{I}  
 	 	\end{array}
    \right]
 	 	\langle X\rangle,
    \label{eq:M2gitterman}	 	
 \end{equation}
with
\begin{equation}
\begin{gathered}
{A_4} = \begin{bmatrix}
	 		0 & 0  & -2g_0^{(4)}\\
            0& 0  & 2h_0^{(4)} \\
            h_0^{(4)} &-g_0^{(4)} & 0
	 	\end{bmatrix}, \\
	{C_2} = \begin{bmatrix}
	 		0 & 0 & -\omega^2 \\
            0 & 0 & \omega^2\\
	 		\frac{\omega^2}{2} & -\frac{\omega^2}{2} & 0
	 	\end{bmatrix},  
   \label{eq:matricesbeta2}
\end{gathered}
\end{equation}
% \begin{equation}
%   \frac{\ud}{\ud z}
% 	 	\langle {X}\rangle= 
% 	 	\begin{bmatrix}
% 		0 & 0  & -2g_0^{(4)}  & 0 & 0 & -2\\
% 		0& 0  & 2h_0^{(4)} & 0 & 0 & 2 \\
% 		h_0^{(4)} &-g_0^{(4)} & 0 & 1& -1& 0\\
% 		0 & 0 & -2\sigma^2_2 & -B_2 & 0 &  -2g_0^{(4)}  \\
% 		0 & 0 & 2\sigma^2_2 &  0 & -B_2 & 2h_0^{(4)} \\
% 		\sigma^2_2 & -\sigma^2_2 & 0 & h_0^{(4)} & -g_0^{(4)} & -B_2 \\
% 	\end{bmatrix}
% 	 	\langle {X}\rangle,
% 	 	\label{eq:M2gitterman}	 	
% \end{equation}
 with $g_0^{(4)}\equiv  \bar\beta_4 \frac{\omega^4}{24}$, $h_0^{(4)} = g_0^{(4)} + 2 \gamma P$ and $\mathbf{I}$ is the identify matrix.
Eq.~\eqref{eq:M2gitterman} is, \textit{mutatis mutandis}, identical to Eq.~(21) in \cite{Armaroli2022}. 

The gain for the second moments is {$G_2(\omega)\equiv \mathrm{Re}\,\tilde\lambda/2$, with $\tilde\lambda$} the eigenvalue of largest real part of the $6\times6$ matrix in Eq.~\eqref{eq:M2gitterman}. Recall that, for  deterministic GVD\cite{Garnier2000}, the gain associated to every moment is identical, thus $G$ in Eq.~\eqref{eq:MIGNLSE} can be considered equivalent to $G_2$. By symbolic manipulations, it is easy to verify that, in the limit of $B_2\to 0$, we obtain again the conventional MI gain, \textit{i.e.}, Eq.~\eqref{eq:MIGNLSE} with $ \pm\beta_2^\mathrm{M}$ in lieu of $\bar\beta_2$. 

\subsection{Synchronous fluctuations of GVD and FOD}
\label{sec:singleprocessGVDFOD}

{Let us now consider a zero average FOD, but non-zero oscillations ($\bar\beta_4 = 0$, $\beta_m^M\neq 0$ for both $m=2,4$).}
First, we consider synchronous fluctuations, \textit{i.e.}, a single process defined by a sample $\{z_n\}$ and $\beta_4^\mathrm{M}=\rho \beta_2^\mathrm{M}$, therefore $\delta\!g = \Delta\! \beta_2(z) (\frac{\omega^2}{2}+\rho \frac{\omega^4}{24})$. Following the same procedure and same variable definition of Sec.~\ref{sec:singleprocessGVD}, we obtain
\begin{equation}
  \frac{\ud \langle X\rangle}{\ud z}
	 	= \left[
 	 	     \begin{array}{c|c}
 	 	     A_0 &  C_{24} \\
 	 	    \hline
 	 	    \sigma^2_2  C_{24} &  A_0-B_2 \mathbf{I}  
 	 	\end{array}
    \right]
 	 	\langle X\rangle,
    \label{eq:M2gitterman2}	 	
 \end{equation}
with
\begin{equation}
\begin{gathered}
{A_0} = \begin{bmatrix}
	 		0 & 0& 0\\0& 0& 4\gamma P\\
	 		2\gamma P & 0  & 0
	 	\end{bmatrix}, \\
	{C_{24}} = \begin{bmatrix}
	 		0 & 0 & -\omega^2 -\rho\frac{\omega^4}{12}\\
            0 & 0 & \omega^2 + \rho\frac{\omega^4}{12}\\
	 		\frac{\omega^2}{2} + \rho\frac{\omega^4}{12} & -\frac{\omega^2}{2} -\rho\frac{\omega^4}{12}& 0
	 	\end{bmatrix},  
   \label{eq:matricesbeta24}
\end{gathered}
\end{equation}

Consider the limit $B_2,B_4\to 0$: by symbolic manipulation, it is easy to verify that we obtain again Eq.~\eqref{eq:MIGNLSE} with $\pm\beta_m^\mathrm{M}$ in lieu of $\bar\beta_m$, for $m=2,4$, respectively. Thus, for $\rho>0$ we obtain a single MI sideband, while for $\rho<0$ we obtain also secondary sidelobes.

 In contrast with the analysis of Eq.~\eqref{eq:M2gitterman}, there are different sign combinations that give MI gain, so we expect a combination of different MI sidelobes corresponding to different nonlinear phase-matching conditions to appear in $G_2$, at least in the limit of $\bar L_2,\bar L_4\to \infty$. 

\subsection{Independent fluctuations of GVD and FOD}

{Let us consider again a zero average FOD, but non-zero independent oscillations ($\bar\beta_4 = 0$, $\beta_m^M\neq 0$ for $m=2,4$, $\langle \Delta\beta_2\Delta\beta_4\rangle=0$ for all $z$).}

 The Shapiro-Loginov formula must be generalized, as in \cite{Klyatskin2010book,Burov2016,Armaroli2022}, to two independent random processes, as follows
\begin{equation}
 \langle\Delta\beta_2\Delta\beta_4 \frac{\ud X_i}{\ud z}\rangle = \left(\frac{\ud}{\ud z} + B_{2}+B_{4}\right)\langle\Delta\beta_2\Delta\beta_4X_i\rangle.  
\end{equation}
The derivation of a closed system of ODEs consists of four steps: (i) average directly Eq.~\eqref{eq:MIeqs2}, (ii) multiply each row of Eq.~\eqref{eq:MIeqs2} by $\Delta\beta_2$ and average, (iii) multiply  by $\Delta\beta_4$ and average; (iv) multiply by $\Delta\beta_2\Delta\beta_4$ and average.
If we assume as above that we can factor the variance out if the same process occurs twice inside an angle bracket, take statistical independence into account, and define  $X_{3+i} \equiv \Delta\!\beta_2 X_i$, $X_{6+i} \equiv \Delta\!\beta_4 X_i$, and $X_{9+i} \equiv \Delta\!\beta_2 \Delta\!\beta_4 X_i$, $i=1,2,3$ we obtain
\begin{equation}
  \frac{\ud \langle {X}\rangle}{\ud z}
	 	= 
	 	\left[
	 	\begin{array}{c|c|c|c}
	 	     A_0 & C_2 &  C_4 & \mathbf{0}\\
	 	    \hline
	 	    \sigma^2_2 C_2 &  A_0-B_2 \mathbf{I}  & \mathbf{0} &  C_4 \\
	 	    \hline
	 	    \sigma^2_4 C_4 & \mathbf{0} &   A_0-B_4 \mathbf{I} &  C_2 \\
	 	    \hline
	 	 \mathbf{0} & \sigma^2_4 C_4 & \sigma^2_2C_2&  A_0-(B_2+B_4) \mathbf{I}
	 	\end{array}
	 	\right]
	 	\langle {X}\rangle,
	 	\label{eq:M4gitterman}	 
\end{equation}
with $\mathbf{0}$ the null matrix,
\begin{equation}
\begin{gathered}
	{C_4} = \begin{bmatrix}
	 		0 & 0 & -\frac{\omega^4}{12} \\
            0 & 0 & \frac{\omega^4}{12}\\
	 		\frac{\omega^4}{24} & -\frac{\omega^4}{24} & 0
	 	\end{bmatrix};  
\end{gathered}
   \label{eq:matricesbeta2beta4}
\end{equation}
 the matrices $A_0$
 and $C_2$ are as in Eqs.~\eqref{eq:matricesbeta2} and \eqref{eq:matricesbeta24},  {and} $\sigma^2_4 \equiv ({\beta_4^\mathrm{M}})^2$. The MI gain is now given by the eigenvalue of largest real part of the $12\times 12$ matrix in Eq.~\eqref{eq:M4gitterman}. 

We consider the limit $B_2,B_4\to 0$. By symbolic manipulation, it is easy to verify that we obtain again Eq.~\eqref{eq:MIGNLSE} with $\pm\beta_m^\mathrm{M}$ in lieu of $\bar\beta_m$, for $m=2,4$, respectively. 

A multi-sidelobe MI gain spectrum is expected as above in Sec.~\ref{sec:singleprocessGVDFOD}, because every possible combination of GVD and FOD signs occurs in any realization.

% In contrast with the analysis of Eq.~\eqref{eq:M2gitterman}, there are different sign combinations that give MI gain, so we expect a combination of different MI sidelobes corresponding to different nonlinear phase-matching conditions to appear in $G_2$, at least in the limit of $\bar L_2,\bar L_4\to \infty$. 

The comparison to numerical results in the next section will show the correctness of this estimates.

\section{Numerical results}
\label{sec:results}
% generalities
In order to validate our analytical results, we generate an ensemble of $N_s=1\times 10^6$ different GVD (and FOD if applies) profiles and solve equation Eq.~\eqref{eq:MIeqs1} exactly for each realization by means of the transfer matrix method\cite{Agrawal2012}. The initial sign of GVD (FOD) is randomized.
Telegraph processes are generated according to the method described in Sec.~\ref{sec:model}.
For the sake of definiteness, { we take $\gamma P = 1$ and ${ \beta_2^\mathrm{M}}=1$, which is equivalent to defining a nonlinear length  $z_\mathrm{nl}\equiv \left(\gamma P\right)^{-1}$ and a characteristic time $t_0\equiv \sqrt{{\beta_2^\mathrm{M}} z_\mathrm{nl}}$ and to  normalizing the propagation distance $ z/z_\mathrm{nl}\to z$, time $t/t_{0}\to t$, field $ U/\sqrt{P}\to U$, and FOD $\beta_4^\mathrm{M}/((\beta_2^\mathrm{M})^2 z_\mathrm{nl})\to \beta_4^\mathrm{M}$.} 
%\textcolor{red}{(COMMENT:\color{red} $\beta_2^\mathrm{M}$ in the square bracket should be the \sout{non}-dimensional quantity? We should add the definition of the normalised FOD coefficient.)} {\color{green} ANSWER: we used the same phrasing in the previous paper. Once everything is normalized, $\beta_4$ is normalized too} 

The domain length is set to $L=20$, initial conditions are chosen to be $(x_1(0),x_2(0))^T=(1,0)$. For each realization, at the end of the---now deterministic---propagation, we compute $P_\mathrm{out}= x_1^2(L) + x_2^2(L)$. The mean gain is defined as
\begin{equation}
    \overline G_2(\omega;N)  \equiv \frac{1}{2  L  }\ln \left\langle\frac{P_\mathrm{out}}{P_\mathrm{in}}\right\rangle,\label{eq:meangain2}
\end{equation}
where the average is performed on the ensemble and has to be compared to $G_2$.

\subsection{Fluctuations of GVD only, $\bar\beta_4=0$ }
As a first example, we consider the effect of an ideal random DM {without any FOD contribution}, \textit{i.e.}, ${\beta_4^\mathrm{M}}=\bar\beta_4=0$.
%%%%%%%%%%%%%%%%%%%%%%%%%%%%%%%%%%%%%%
% Figure: two examples of telegraph GVD w/ different Lav
\begin{figure}[!ht]
    \centering
    \includegraphics[width=0.5\textwidth]{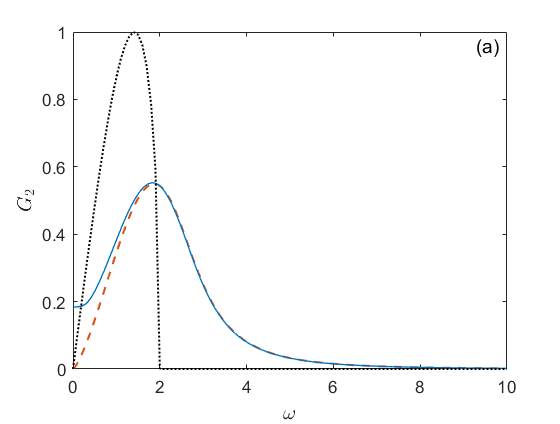}
     \includegraphics[width=0.5\textwidth]{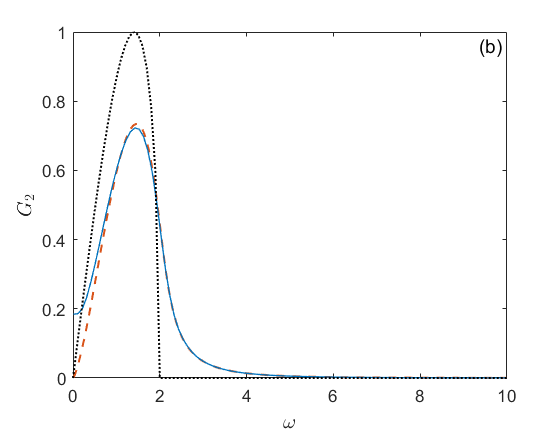}
    \caption{MI gain as a function of detuning $\omega$ for a random GVD distribution with two different average lengths (a) $\bar L_2 = 0.2$, (b) $\bar L_2 = 1.2$. The numerical results (solid blue curve) are compared to the numerical estimate of Eq.~\eqref{eq:M2gitterman} (red dahsed curves) and to the conventional MI gain in a uniform anomalous GVD fiber with $\bar\beta_2=-1$ (black dotted curve).}
    \label{fig:telegraphGVD}
\end{figure}
In Fig.~\ref{fig:telegraphGVD}(a) we show a quite short mean length $\bar L_2 = 0.2$ (large $B_2=10$), comparable to the values studied in \cite{Armaroli2022}. The MI gain consists of one single sidelobe centered around $\omega\approx 2$ and much broader than the conventional MI in anomalous GVD (shown as a dotted black line for comparison), with a maximum about 50\% of it. The very large ensemble yields a very smooth $\overline G_2$ (solid blue line) that
 fits almost perfectly to the corresponding $G_2$ (dashed red line), apart from small $\omega$, due to finite size effects, as explained at length in \cite{Armaroli2022,Armaroli2023DM}. If a larger length $\bar L_2 = 1.2$ is used, see Fig.~\ref{fig:telegraphGVD}(b), the MI sidelobe resembles to the conventional MI one, the maximum occurs around $\omega=\sqrt{2}$, but for a smooth tail beyond the convectional cutoff at $\omega=2$ and a smaller maximum value (about 70\% of the convetional one).

%%%%%%%%%%%%%%%%%%%%%%%%%%%%%%%%%%%%%%
% Figure: two examples of telegraph GVD w/ different Lav
\begin{figure}[!h]
    \centering
    \includegraphics[width=0.5\textwidth]{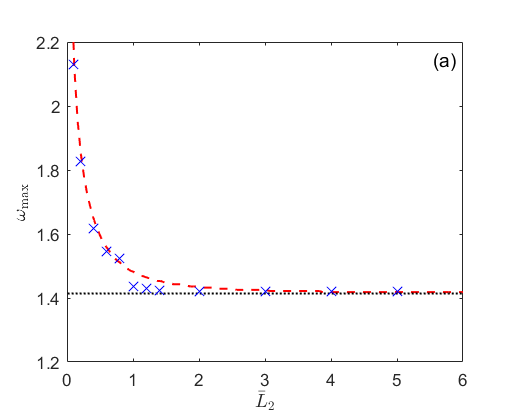}
     \includegraphics[width=0.5\textwidth]{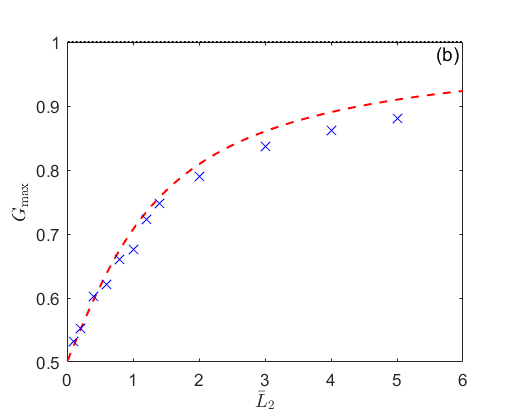}
    \caption{Comparison of 
maxima of $G_2$ and $\overline G_2$ as a function $\bar L_2$: (a) the maximum points (b) the maximal values. Dashed red lines correpond to the analytical estimates, blue crosses to numerical results. The black dotted line in (a) represents the conventional MI value $\omega=\sqrt{2}$. In panel (b) this limit is $G_2=1$. }
    \label{fig:telegraphGVDmax}
\end{figure}

To thoroughly assess this properties, we show in Fig.~\ref{fig:telegraphGVDmax} the maximum point $\omega_\mathrm{max}$ (a) and its value $G_\mathrm{max}$ (b) as a function of $\bar L_2$. We notice that both converge for $\bar L_2\to \infty$ to the conventional MI values. The convergence of $\omega_\mathrm{max}$ is much faster than that of $G_\mathrm{max}$. The discrepancies of this latter at large $\bar L_2$ depend on the limited domain length.

\subsection{Fluctuations of GVD only, $\bar\beta_4\neq 0$ }

As a second example, we consider the impact of {constant} FOD on the ideal random DM, \textit{i.e.}, ${ \beta_4^\mathrm{M}}= 0$ but $\bar\beta_4\neq 0$. We take $\bar L_2=1.2$, as in the example reported in Fig.~\ref{fig:telegraphGVD}(b).

%%%%%%%%%%%%%%%%%%%%%%%%%%%%%%%%%%%%%%%%%%%
% Figure: impact of FOD at fixed Lav (telegraph GVD)
\begin{figure}[!h]
    \centering
    \includegraphics[width=0.5\textwidth]{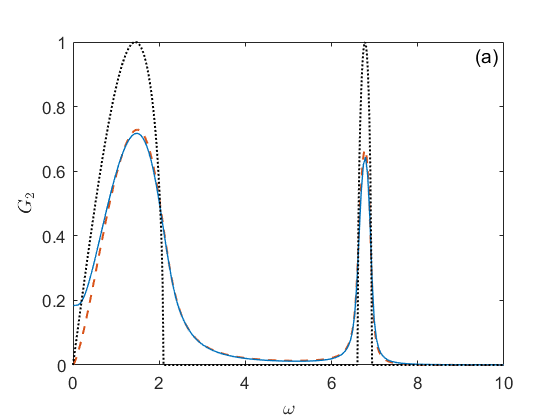}
     \includegraphics[width=0.5\textwidth]{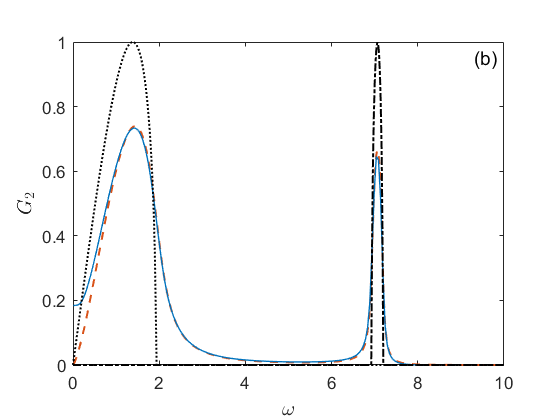}
    \caption{MI gain as a function of detuning $\omega$ for a random GVD distribution with intermediate $\bar L_2 = 1.2$ and fixed FOD. { As in Fig.~\ref{fig:telegraphGVD}, the numerical results (solid blue curve) are compared to the numerical estimate of Eq.~\eqref{eq:M2gitterman} (red dahsed curves). The dotted lines show the MI gain in a uniform anomalous GVD fiber with $\bar\beta_2=-1$ and positive (negative) FOD in (a) [resp.~ (b)], while the additional dash-dotted line in panel (b) shows the  MI gain in a uniform normal GVD fiber with $\bar\beta_2=1$ and negative FOD.  Chosen values of FOD are (a) ${\overline\beta_4} = 0.25$,  (b) ${\overline\beta_4} = -0.25$.}}
    \label{fig:telegraphGVDplusfixedFOD}
\end{figure}
We show in Fig.~\ref{fig:telegraphGVDplusfixedFOD} the impact of a positive, $\bar\beta_4=0.25$ in panel (a), or negative, $\bar\beta_4=-0.25$ in panel (b), FOD. As expected, beyond the conventional low-frequency sidelobe, which stems from the anomalous GVD segments, the nonlinear phase-matching condition in Eq.~\eqref{eq:MIpeaks} leads to the appearance of an additional MI sidelobe. 
% panel (a) positive FOD
In Fig.~\ref{fig:telegraphGVDplusfixedFOD}(a) this correspond to the position of the {high frequency} sidelobe {predicted by} Eq.~\eqref{eq:MIGNLSE}, by  substituting $\bar\beta_2\to-{\beta_2^\mathrm{M}} $, \textit{i.e.}, the  phase-matching by the segments of anomalous GVD gives rise to a second sidelobe at $\omega\approx6.77$, which achieves a smaller value of $G_2$ and is broader than in a homogeneous fiber. Moreover, the primary  (baseband) sidelobe attains a larger gain than the secondary one, at variance with the homogeneous counterpart (compare the blue solid  to the dotted black line). 
%panel (b) negative FOD
A similar effect is observed in  Fig.~\ref{fig:telegraphGVDplusfixedFOD}(b), where by substituting $\bar\beta_2\to{ \beta_2^\mathrm{M}}$ in Eq.~\eqref{eq:MIGNLSE}, we can explain the secondary sidelobe appearing at $\omega\approx7.06$ (dash-dotted line). The impact of the different FOD sign on the primary MI sidelobe  and on the imbalance between the two is negligible. 

% discussioon of figures
\subsection{Synchronous  fluctuations of  GVD and FOD, $\bar\beta_4=0$}

As a third example, we let both GVD and FOD vary  around a null mean value, $\bar\beta_2=\bar\beta_4=0$, in a synchronous way. We take $\rho = -0.25$ to facilitate the comparison with the previous example in Fig.~\ref{fig:telegraphGVDplusfixedFOD}. 
\begin{figure}[!h]
    \centering
    \includegraphics[width=0.5\textwidth]{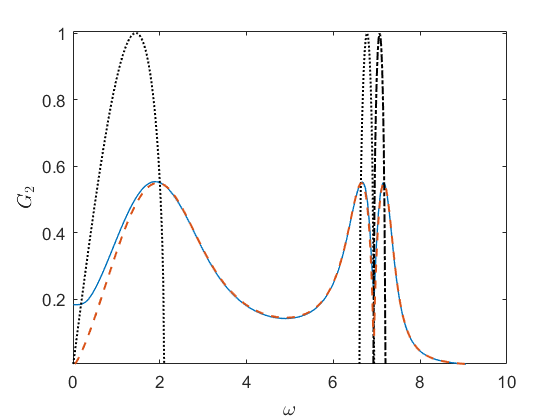}
     \includegraphics[width=0.5\textwidth]{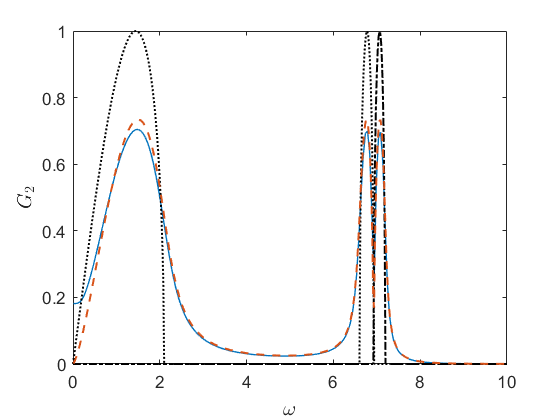}
    \caption{Same  as in 
Fig.~\ref{fig:telegraphGVDplusfixedFOD} for random and synchronous GVD and FOD with (a) $\bar L = 0.2$, (b) $\bar L = 1.2$. {The dotted lines correspond here to the gain in a homogeneous fiber with $\bar\beta_2 = -1$, $\bar\beta_4 = 0.25$, the dash-dotted lines to $\bar\beta_2 = 1$, $\bar\beta_4 = -0.25$. The MI sidelobe $\bar\beta_2 = -1$, $\bar\beta_4 = -0.25$ is almost identical to the conventional MI shown in Fig.~\ref{fig:telegraphGVD} and is omitted.}}
    \label{fig:telegraphGVDandFOD_S}
\end{figure}

We consider two different mean lengths $\bar L=\{0.2,1.2\}$. Therefore, Fig.~\ref{fig:telegraphGVDandFOD_S} is compared directly to Fig.~\ref{fig:telegraphGVD}. For both mean lengths, the primary sidelobe at small $\omega$ is indistinguishable if FOD is present or not, like in Fig.~\ref{fig:telegraphGVDplusfixedFOD}. A pair of secondary sidelobes appear and turn out to exhibit nearly the same gain of the primary, around 50$\%$ and 70$\%$ of the conventional one, respectively, as shown above. This is consistent with the discussion in Sec.~\ref{sec:model}: the sidelobe at $\omega\approx 6.77$  (along with the primary one, see dotted black line) originates from anomalous GVD segments predicted by Eq.~\eqref{eq:MIGNLSE} by substituting $\bar\beta_2\to -\beta_2^\mathrm{M}$, the sidelobe at $\omega\approx 7.06$  (dash-dotted black line) from normal GVD segments, as predicted  by Eq.~\eqref{eq:MIGNLSE} by substituting $\bar\beta_2\to \beta_2^\mathrm{M}$. Each segment in the link gives thus rise to an independent phase matching condition and the corresponding MI sidelobe. The choice of $\rho\ge 0 $ (not shown) gives obviously only the primary one. As in Fig.~\ref{fig:telegraphGVD}, the sidelobes are broader for smaller $\bar L_2$. 

\subsection{Independent fluctuations of  GVD and FOD, $\bar\beta_4=0$}

We finally let both GVD and FOD vary independently around a null mean value, $\bar\beta_2=\bar\beta_4=0$. We take ${\beta_4^\mathrm{M}} = 0.25$ to facilitate the comparison with the previous example in Fig.~\ref{fig:telegraphGVDandFOD_S}. We assume $\bar L _2 = \bar L_4 \equiv \bar L$.
%%%%%%%%%%%%%%%%%%%%%%%%%%%%%%%%%%%%%%%%
% Telegraph processes for both GVD and FOD
\begin{figure}[!h]
    \centering
    \includegraphics[width=0.5\textwidth]{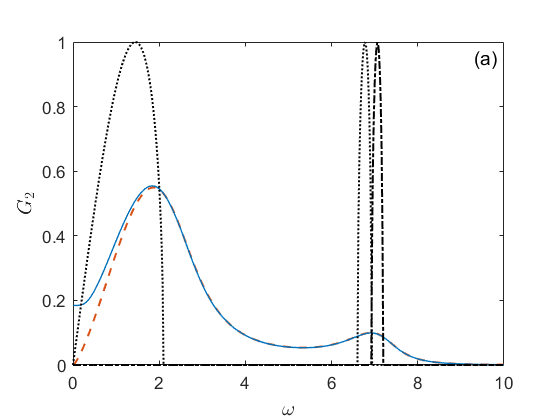}
     \includegraphics[width=0.5\textwidth]{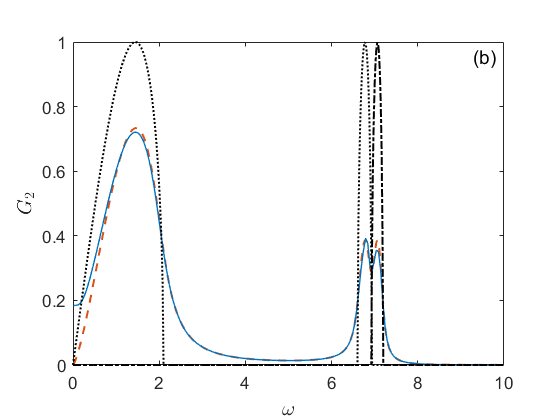}
    \caption{Same  as in Fig.~\ref{fig:telegraphGVDandFOD_S} for random and independent GVD and FOD with (a) $\bar L = 0.2$, (b) $\bar L = 1.2$. }
    \label{fig:telegraphGVDandFOD_I}
\end{figure}
We consider again two different mean lengths $\bar L=\{0.2,1.2\}$. Therefore, Fig.~\ref{fig:telegraphGVDandFOD_I} is compared directly to Fig.~\ref{fig:telegraphGVDandFOD_S}. The primary sidelobe at small $\omega$ is again indistinguishable if FOD is present or not, apart from the tails connecting it to the secondary sidelobes (in the $3<\omega< 5$ range). These latter, stemming from the fluctuations of FOD, exhibit instead several peculiar properties, quite different from those found for synchronous fluctuations. For $\bar L= 0.2$, a small sidelobe occurs at $\omega\approx 6.92$ and is about 6 times smaller than the first one. For larger $\bar L = 1.2$ the secondary sidelobe  splits into two---we verify that they start to be distinughible at $\bar L =0.8$ (not shown). By comparing to the MI gain given by every  possible homogeneous limit, \textit{i.e.}, every combination of $(\bar\beta_2\to \pm{ \beta_2^\mathrm{M}}, \bar\beta_4\to \pm{ \beta_4^\mathrm{M}})$---dotted (dash-dotted) black lines for $\bar\beta_2< 0$ ($\bar\beta_2>0$), respectively---it is apparent that the pair of secondary sidelobes correspond to the same nonlinear phase-matching conditions discussed above in reference to Fig.~\ref{fig:telegraphGVDplusfixedFOD}(a) and \ref{fig:telegraphGVDplusfixedFOD}(b), respectively, and to Fig.~\ref{fig:telegraphGVDandFOD_S}. The independent oscillations of GVD and FOD give rise, effectively, to all the possible phase-matching conditions being satisfied and thus to three independent maxima of $G_2$. The secondary lobes achieve a smaller $G_2$ in Fig.~\ref{fig:telegraphGVDandFOD_I}(b)  (54\% of the value of the primary one) than in Figs.~\ref{fig:telegraphGVDplusfixedFOD} and \ref{fig:telegraphGVDandFOD_S}(b), on account of the FOD stochasticity. As discussed in \cite{Armaroli2022}, larger $\omega$ are more sensitive to lowpass random fluctuations, because the associated wavelength is smaller and thus comparable to the mean length. Combining random fluctuations on two parameters amplifies this effect.

\section{Conclusions}
\label{sec:concl}

% short comment on the nonlinear behavior beyond the linearized approximation. 
We studied a specific (lowpass) colored stochastic process, the telegraph process, as a fluctuation of group-velocity dispersion around a zero mean value. This resembles a random dispersion-manged fiber link. We show how the modulational instability sidebands behave as a function of the mean \textit{waiting} length between sign changes: for a short mean length the MI sidelobes are broader and smaller than their conventional counterpart, but they converge to them for a long mean length (several times the nonlinear length). We included also, as an additional refinement, the effect of a constant or variable FOD. A constant FOD (pertinent if the operation point is close to zero-dispersion point) yields an additional sideband, centered around an easily predictable detuning. { Fluctuating FOD may give rise to a pair of closely separated secondary lobes, which correspond to the set of all the sidelobes in normal or anomalous GVD, with different FOD signs. Similarly to the MI gain in a homogenous fiber, where even-order ($2n$) high-order dispersion yields a family of up to $n$ sidelobes of equal maximum gain, the stochastic counterpart shows balanced sidelobes for synchronous fluctuations. This is not the case for GVD variations around constant FOD or independent GVD and FOD variations, where  the baseband  is larger than the high-frequency sidelobes. For short correlation lengths, a single merged high-frequency sidelobe appears for independent GVD and FOD variations.} In every example, the analytical estimates obtained by the  analytical approach we employ here (after M.~Gitterman's  work) match almost perfectly with the numerical data.

A few other combinations of GVD and FOD could be analyzed, but for the sake of brevity we decided to focus on just {four} of them, which we find of more practical interest. A pure FOD fluctuation will yield a negligible MI gain, as in Ref.~\cite{Armaroli2022} for pure GVD, if $\bar\beta_4 \neq 0$ or else converge to the FOD gain profile\cite{Cavalcanti1991}, \textit{i.e.}, a quadratic growth for small $\omega$. Fluctuations of GVD around $\bar\beta_2=0$, associated to $\bar\beta_4\neq 0$ and ${\beta_4^\mathrm{M}}\neq 0 $ yield a proliferation of MI sidebands, satisfying all the admissible nonlinear phase-matching conditions as explained in the present work.   

%\textcolor{red}{COMMENT: I do not really understand this paragraph. I don't think it is necessary, maybe we can leave if as an answer to a possible referee's question? {\color{blue} ANSWER: is it dangerous? If not we can keep it. In case, we cross the same people making the same remarks.}
%We limited ourselves to the linearized NLSE. Several studies are devoted to nonlinear oscillators with multiplicative noise, see for example \cite{Mallick2002,Mallick2003}. \sout{It is well-known that the NLSE can be reduced to an oscillator with a quartic potential. Nevertheless, as sidebands grow, the amount of energy they can acquire is limited by $P$, \textit{i.e.}, the potential is not of the type studied in those papers. It is thus quite complicate to formulate a Fokker-Planck equation for characterizing the scaling of diffusion. Qualitatively, we expect that, after the exponential growth and a short slowdown stage of possibly anomalous diffusion, the average orbit will settle to encircle the origin of the $(x_1,x_2)$ plane with no specific symmetry. } This will, nonetheless, be the object of a future study.}
{We limited ourselves to the linearized NLSE, which models the initial growth of unstable sidebands. The nonlinear stage which takes place as soon as enough energy is converted from the pump $P$ to sidebands is not considered here. Several studies are devoted to nonlinear oscillators with multiplicative noise, see for example \cite{Mallick2002,Mallick2003}. Their application to optical fibers will be the object of a future study.}

Our results may permit to  tailor MI gain sidebands in
optical fibers by means of stochastic GVD fluctuations and suggest the regimes to achieve that. For example, 
such discrete fluctuations can be implemented by means of fiber splicing.

\begin{backmatter}
\bmsection{Funding}
The present research was supported by IRCICA (USR 3380 CNRS), Agence Nationale de la Recherche (Programme Investissements d’Avenir, I-SITE VERIFICO, Labex CEMPI); Ministry of Higher Education and Research; Hauts de France Council; European Regional Development Fund (Photonics for Society P4S, Wavetech); Italian Ministry of Higher Education and Research, PRIN grants (2020X4T57A, 20222NCTCY)

\bmsection{Acknowledgments}
We thank S.~De Bi{è}vre and G.~Dujardin of Laboratoire Paul Painlevé for fruitful discussions. 

\bmsection{Disclosures}
The authors declare no conflicts of interest.

\bmsection{Data availability} Data underlying the results presented in this paper are not publicly available at this time but may be obtained from the authors upon reasonable request.

\end{backmatter}

%%%%%%%%%%%%%%%%%%%%%%% References %%%%%%%%%%%%%%%%%%%%%%%%%

%%%%%%%%%% If using BibTeX:
\bibliography{RandomFibers}

%%%%%%%%%% If preparing manually:
% \begin{thebibliography}{1}
% \newcommand{\enquote}[1]{``#1''}

% \bibitem{Zhang:14}
% Y.~Zhang, S.~Qiao, L.~Sun, Q.~W. Shi, W.~Huang, L.~Li, and Z.~Yang,
%   \enquote{Photoinduced active terahertz metamaterials with nanostructured
%   vanadium dioxide film deposited by sol-gel method,}
%   {\protect\JournalTitle{Optics Express}} \textbf{22}, 11070--11078 (2014).

% \bibitem{Optica}
% {Optica}, \enquote{{Optica Publishing Group},}
%   \url{http://www.opg.optica.org}.

% \bibitem{FORSTER2007}
% P.~Forster, V.~Ramaswamy, P.~Artaxo, T.~Bernsten, R.~Betts, D.~Fahey,
%   J.~Haywood, J.~Lean, D.~Lowe, G.~Myhre, J.~Nganga, R.~Prinn, G.~Raga,
%   M.~Schulz, and R.~V. Dorland, \enquote{Changes in atmospheric consituents and
%   in radiative forcing,} in \enquote{Climate Change 2007: The Physical Science
%   Basis. Contribution of Working Group 1 to the Fourth assesment report of
%   Intergovernmental Panel on Climate Change,}  S.~Solomon, D.~Qin, M.~Manning,
%   Z.~Chen, M.~Marquis, K.~B. Averyt, M.~Tignor, and H.~L. Miler, eds.
%   (Cambridge University Press, 2007).

% \end{thebibliography}

\end{document}